\documentclass[onecolumn,showpacs,aps,epsfig,nofootinbib]{revtex4}

\usepackage{graphicx}
\usepackage{epstopdf}
\usepackage{latexsym}
\usepackage{graphicx,times}
\usepackage{multirow}
\usepackage{natbib}
\usepackage{amssymb,amsmath}
\usepackage{color}
\usepackage{colordvi}
\usepackage{mathrsfs}

\usepackage[center]{subfigure}

\begin{document}

 \newcommand{\bq}{\begin{equation}}
 \newcommand{\eq}{\end{equation}}
 \newcommand{\bqn}{\begin{eqnarray}}
 \newcommand{\eqn}{\end{eqnarray}}
 \newcommand{\nb}{\nonumber}
 \newcommand{\lb}{\label}
 \newcommand{\tc}{\textcolor{black}}
\newcommand{\PRL}{Phys. Rev. Lett.}
\newcommand{\PL}{Phys. Lett.}
\newcommand{\PR}{Phys. Rev.}
\newcommand{\CQG}{Class. Quantum Grav.}

\title{Thermal gravitational-wave background in the general pre-inflationary scenario}

\author{Kai Wang$^1$, Larissa Santos$^1$, Jun-Qing Xia$^2$, Wen Zhao$^1$\footnote{Corresponding author. \\ E-mail address: wzhao7@ustc.edu.cn (W. Zhao)}}
 \affiliation{$^1$ CAS Key Laboratory for Researches in Galaxies and Cosmology, Department of Astronomy, University of Science and Technology of China, Chinese Academy of Sciences, Hefei, Anhui 230026, China \\
 $^2$ Department of Astronomy, Beijing Normal University, Beijing 100875, China}

\date{\today}

\begin{abstract}

We investigate the primordial gravitational waves (PGWs) in the
general scenario where the inflation is preceded by a
pre-inflationary stage with the effective equation of state $w$.
Comparing with the results in the usual inflationary models, the
power spectrum of PGWs is modified in two aspects: One is the
mixture of the perturbation modes caused by he presence of the
pre-inflationary period, and the other is the thermal initial
state formed at the Planck era of the early Universe. By
investigating the observational imprints of these modifications on
the B-mode polarization of cosmic microwave background (CMB)
radiation, we obtain the constraints on the conformal temperature
of the thermal gravitational-wave background $T<5.01\times
10^{-4}$Mpc$^{-1}$ and a tensor-to-scalar ratio $r<0.084$
($95\%$ confident level), which follows the bounds on total number
of e-folds $N>63.5$ for the model with $w=1/3$, and $N>65.7$ for
that with $w=1$. By taking into account various noises and the
foreground radiations, we forecast the detection possibility of
the thermal gravitational-wave background by the future CMBPol
mission, and find that if $r>0.01$, the detection is possible as
long as $T>1.5\times 10^{-4}$Mpc$^{-1}$. However, the effect of
different $w$ is quite small, and it seems impossible to determine
its value from the potential observations of CMBPol mission.

\end{abstract}


\maketitle

\section{Introduction}

In modern cosmology, the early Universe is a very attractive topic, which provides an excellent experimental opportunity to study the physics in an extremely high energy and strong gravitational field. It may also give us a glimpse at the physical conditions of the very early Universe right to the time of its birth \cite{earlyUniverse}. In the standard inflationary scenario, the early Universe had a nearly de Sitter expansion. During this stage, the nearly scale-invariant primordial power spectra of density perturbations (i.e. scalar perturbations) and gravitational waves (i.e. tensor perturbations) originated from the zero-point quantum fluctuations \cite{grishchuk1979} successfully explains various observations on cosmic microwave background (CMB) radiation \cite{cmb} and the distributions of
the galaxies in various large-scale structure observations.

In the standard calculations of the primordial perturbations in
the inflationary models, the initial condition is always chosen as
the Bunch-Davies vacuum (see for instance, \cite{Baumann:2009ds}).
However, if considering the pre-inflationary stage
\cite{pre-inflation}, this vacuum choice should be modified in
general, which is determined by the physics and evolution of
Universe in the pre-inflationary stage. Nowadays, with the
observational improvement of the CMB temperature and polarization
fluctuations, it becomes possible to carefully test the initial
condition of the primordial power spectra, which might leave
significant imprints on the CMB fluctuations on the largest
scales.

The need of a pre-inflationary stage can also be understood by the following way: it seems logical to suggest that our Universe came
into being as a configuration with a Planck size and a Planck energy density, and with a total energy, including
gravity, equal to zero (see \cite{zeldovich} and references therein). The newly created classical configuration cannot directly reach the
average energy density of inflationary stage, which is around at or lower than the Grand unification energy scale \cite{zeldovich}. In order to connect the initial Planck state and the inflationary stage, a radiation dominant period is always considered in the previous works \cite{w13,bha,zhao2009,das}. However, the physics of the pre-inflationary stage is unclear for us. It is possible that, a scalar field might dominate the evolution of the Universe during this stage. If the scalar field was dominant by its kinetic energy, the effective equation of state (EoS) was $w=1$, instead of $w=1/3$ \cite{zhutao}. On the other hand, if the pre-inflationary stage was dominant by the matter component, one has $w=0$. In order to avoid this uncertainties, in this paper, we consider a general scenario for the pre-inflationary stage, in which the effective EoS $w$ is a free parameter, and investigate the effect of different $w$ values.

In the pre-inflationary scenario, it is reasonable to assume that
during the stage at temperatures higher than $\sim 10^{19}$GeV a
thermal equilibrium between various components, including
gravitons, is maintained through gravitational interaction
\cite{earlyUniverse,zhao2009,das,zhutao}. As the Universe cooled down and the
gravitons decoupled, a background of thermal relic gravitons with
a black-body spectrum would be left behind. The following
evolution of the thermal background of gravitational waves depends
only on the evolution of the cosmic scale factor
\cite{gw-evolution}. Thus, the detection of this background gives
a unique chance to probe the physics of pre-inflationary Universe,
since it is inaccessible by other means
\cite{w13,bha,zhao2009,das}. In the previous work \cite{zhao2009},
we have detailed studied the imprints of thermal
gravitational-wave background on the power spectra of primordial
gravitational waves (PGWs) and CMB temperature and polarization
fluctuations, and derived the constraints on the conformal
temperature of the thermal gravitational-wave background. In this
letter, we shall extend this investigation to the general
pre-inflationary scenario by assuming the effect EoS $w$ being a
free parameter.

\section{Thermal background of primordial gravitational waves in the general pre-inflationary scenario}
In the standard slow-roll inflationary scenario, PGWs are thought
to be generated from vacuum fluctuations at the beginning of  the
inflationary stage. These quantum fluctuations will be converted
into classical perturbations by stretching out the particle
horizon, i.e. a decoherence process. Then the perturbation will
re-enter the horizon, where the power spectrum is a nearly
scale-invariant form in this standard scenario
\cite{grishchuk1979,gw-evolution}. However, if considering the
pre-inflationary stage, the choice of the initial state will be
different from the vacuum. In this scenario, there are two main
differences with respect to the standard scenario mentioned above.
Firstly, we consider that the gravitational waves were produced by
the decoupling of gravitons with other particles, which might be
in a black-body distribution \cite{bha}. Secondly, we consider a
pre-inflationary stage to connect the Planck era and the
inflationary stage. The evolution of PGWs during this period will
leave significant imprints in the current power spectrum, which
provides a window to probe the extreme early Universe. In the
following subsections, we will demonstrate the evolution of both
the background and the gravitational waves in the pre-inflationary
scenario, and get the compact form of the power spectrum of PGWs.

\subsection{Expansion stages of the Universe}
In the scenario where the Universe was born from a Planck state,
its energy scale is expected to be the Planck energy scale
\cite{zeldovich}. At the same time, we notice that the energy
scale of inflation is always thought to be below the Planck energy
scale in inflationary models \cite{Baumann:2009ds}. As a result,
there should be a pre-inflationary stage, which acts as a bridge
between the Planck initial stage and the inflationary stage. The
characters and the evolution of this period are determined by the
components and their microscopic physics in the Universe, which is
unknown for us. However, in this paper, we shall only focus on the
PGWs, which depends only on the evolution information of the scale
factor $a$. As a general consideration, we parameterize this stage
by a simple EoS parameter, $w=p/\rho$, i.e. we assume that in the
pre-inflationary period, the effective EoS of the dominant
component in the Universe is a constant (which can also be
understood as the average EoS during this stage). In many papers,
it is always assumed that this stage is radiation dominant, which
follows that $w=1/3$ \cite{w13}. However, in some other models, it
is argued that this stage may be dominant by the scalar field
\cite{zhutao} or some other components. In this paper, we avoid
this kind of uncertainties by assuming $w$ as a free parameter.

By applying the Friedmann equations, we can get the background
evolution of the inflationary and pre-inflationary stages as
follows \cite{Zhang:2005nw,Zhang:2006mja},
{\color{black}\begin{equation}\label{a-evolution}
\begin{cases}
a(\tau)=(\tau/\tau_i) ^{1+\alpha}     &  \tau <\tau_1\\
a(\tau)=[-(\tau-\tau_{\text{inf}})/\tau_0]^{1+\beta}       & \tau>\tau_1,
\end{cases}
\end{equation}}
where $\tau$ is the conformal time, which relates to the cosmic time $t$ by $dt=ad\tau$. $\tau_1$ is the conformal time of transfer from pre-inflationary stage to the inflationary stage. The index $\alpha$ is determined by $\alpha = \frac{1-3w}{1+3w}$. The value of $\beta$ is determined by the inflationary models. For exactly the de Sitter expansion models of inflation, we have $\beta=-2$. While, for the slow-roll models, the value of $\beta$ is slightly larger than $-2$. 
{\color{black}$\tau_i$, $\tau_0$ and $\tau_{\text{inf}}$ are constants, and the values of $\tau_0$ and $\tau_{\text{inf}}$ are fixed by the continue conditions,}

\begin{eqnarray}
a(\tau_1)|_{\text{pre-inflation}}&=&a(\tau_1)|_{\text{inflation}},~~~~\\
a'(\tau_1)|_{\text{pre-inflation}}&=&a'(\tau_1)|_{\text{inflation}}.
\end{eqnarray}
where the \emph{prime} denotes $d/d\tau$ {\color{black} and
$\tau_i$ can be given by the normalization of scale factor}. It
should be note that the pre-inflation started at
$\tau=\tau_{\text{Pl}}>0$ in this parameterization. In addition,
Universe will exit the inflationary stage when
$\tau<\tau_{\text{inf}}$ with $a(\tau)=a_p$, where $a_p$ is the
scalar factor at the beginning of the post-inflationary stage.



\subsection{Evolution of gravitational waves}
Incorporating the perturbations to the spatially flat Friedmann-Lemaitre-Robertson-Walker spacetime, the metric is
\begin{equation}
ds^2=a(\tau)^2[-d\tau ^2 + (\delta_{ij}+h_{ij}(\tau))dx^idx^j],
\end{equation}
where the perturbations of space-time $h_{ij}$ is a $3\times 3$ symmetric matrix. The gravitational-wave field is the tensorial portion of $h_{ij}$, satisfying the transverse-traceless conditions $h_{ij,j}=0$, $h^i_i=0$, which can be decomposed as the $+$ and $\times$ polarization component. As in general, we can expand $h_{ij}$ in Fourier space as follows,
\begin{equation}
h_{ij}(\tau)=\int \frac{d^3k}{(2\pi)^3}\sum_{s=+,\times}\epsilon_{ij}^{s}(k)h^s_{\bf k}(\tau)e^{i{\bf k \cdot x}},
\end{equation}
where $k$ is the wavenumber, and the polarization tensor $\epsilon_{ij}$ satisfies $\epsilon_{ii}=k^i\epsilon_{ij}=0$ and $\epsilon_{ij}^s\epsilon_{ij}^{s'}=2\delta_{ss'}$.

In order to calculate the primordial gravitational waves, it is convenient to define the canonically normalized field $v_{\bf k}^s\equiv\frac{a}{2}M_{\rm Pl}h_{\bf k}^s$, where $M_{\rm Pl}$ is the Planck energy scale. The quantization of the field $v$ is straightforward. The Fourier components $v_{\bf k}^s$ are promoted to operators and expressed via the following decomposition (see for instance, \cite{Baumann:2009ds}),
 \begin{equation}
 v_{\bf k}^s\rightarrow \hat v^s_{\bf k}(\tau)=v^s_{k}(\tau)\hat a_{\bf k}+{v^s_{-k}}^{*}(\tau)\hat a_{-\bf k}^{\dagger},
 \end{equation}
where the function $v^s_k$ satisfies the evolution equation
 \begin{equation}\label{evolution-eq}
 {v^s_{k}}''+(k^2-\frac{a''}{a})v^s_{k}=0.
 \end{equation}
The creation and annihilation operators $a_{-\bf k}^{\dagger}$ and
$a_{\bf k}$ satisfy the canonical commutation relation $[\hat a_{\bf k},\hat a_{\bf k'}^{\dagger}]=(2\pi)^3\delta^3 ({\bf k}-{\bf k'})$ if and only if the mode functions are normalized as,
\begin{equation}\label{normalization}
\frac{i}{\hbar}({v^s_{k}}^*{v}^s_{k}{'}-{{v}^s_{k}}^{*}{'}v^s_{k})=1,
\end{equation}
which provides one of the boundary conditions on the solutions of
Eq. (\ref{evolution-eq}).

The general solutions of Eq. (\ref{evolution-eq}) in pre-inflationary and inflationary stages are given by \cite{Baumann:2009ds,Riotto:2002yw}
\begin{equation}\label{eq9}
\begin{cases}
v_k(\tau)=\sqrt{\tau}[A_1H_{\alpha+\frac12}^{(1)}(k\tau)+A_2H_{\alpha+\frac12}^{(2)}(k\tau)]     &  \tau<\tau_1 \\
v_k(x)=\sqrt{-x}[B_1H_{\beta+\frac12}^{(1)}(-kx)+B_2H_{\beta+\frac12}^{(2)}(-kx)]       &\tau>\tau_1,
\end{cases}
\end{equation}
where $x\equiv \tau-\tau_{\text{inf}}$. $H_{\nu}^{(1)}(x)$ and $H_{\nu}^{(2)}(x)$ are the Hankel's function of the first and the second kind. Obviously, we have four constants to be specified, $A_1$, $A_2$, $B_1$ and $B_2$.

In order to determine $A_1$ and $A_2$, we need to know the other initial condition in addition to the normalization condition in Eq. (\ref{normalization}). It is reasonable that the spacetime of a small region $(k\to \infty)$ should be described by the Minkowski metric. In addition, we have $\tau > \tau_{\text{Pl}} > 0$, where $\tau_{\text{Pl}}$ corresponds to the time when the Universe is at Planck energy scale. Thus, the asymptotic behavior of the mode function $v_k$ should be \cite{Baumann:2009ds}
\begin{equation}\label{initial}
\lim\limits_{k\tau\to \infty}v_k=\frac{1}{\sqrt{2k}}e^{-ik\tau}.
\end{equation}
Considering the two initial conditions, Eq. (\ref{normalization}) and Eq. (\ref{initial}), and the asymptotic property of Hankel's function \cite{handbook}
\begin{equation}
\lim\limits_{x\to \infty}H^{(1,2)}_{\nu}(x)=\sqrt{\frac{2}{\pi x}}e^{\pm i(x-\frac12\nu\pi-\frac14\pi)},
\end{equation}
we get that $A_1=0$ and $A_2=\sqrt{\frac{\pi}{2}}e^{-i\frac{\pi}{2}(\alpha +1)}$, and
\begin{equation}
v_k(\tau)=\sqrt{\frac{\pi \tau}{2}}H_{\alpha+\frac12}^{(2)}(k\tau),~~\tau<\tau_1,
\end{equation}
where we have dropped the phase factor because it will be
cancelled out in the calculation of the power spectrum. We can
determine the other two constants by joining the scale factor and
mode function continuously at $\tau_1$, i.e. \cite{Zhang:2006mja}
\begin{equation}
\begin{cases}
v_k(\tau_1)|_{\text{pre-inflation}}=v_k(\tau_1)|_{\text{inflation}}    &\\
v_k'(\tau_1)|_{\text{pre-inflation}}=v_k'(\tau_1)|_{\text{inflation}}.
\end{cases}
\end{equation}
The results are
\begin{eqnarray}
B_1(k)&=&k\sqrt{\frac{-\pi^3x_1\tau_1}{32}}[H_{\beta+\frac12}^{(2)}(-kx_1)H_{\alpha+\frac32}^{(2)}(k\tau_1)\nonumber
\\ &-&H_{\beta+\frac32}^{(2)}(-kx_1)H_{\alpha+\frac12}^{(2)}(k\tau_1)]\nonumber\\
&-&\sqrt{\frac{-\pi^3(1+\alpha)(1+\beta)}{8}}H_{\beta+\frac12}^{(2)}(-kx_1)H_{\alpha+\frac12}^{(2)}(k\tau_1)\nonumber,
\\B_2(k)&=&k\sqrt{\frac{-\pi^3x_1\tau_1}{32}}[H_{\beta+\frac32}^{(1)}(-kx_1)H_{\alpha+\frac12}^{(2)}(k\tau_1)\nonumber
\\&-&H_{\beta+\frac12}^{(1)}(-kx_1)H_{\alpha+\frac32}^{(2)}(k\tau_1)]\nonumber\\
&+&\sqrt{\frac{-\pi^3(1+\alpha)(1+\beta)}{8}}H_{\beta+\frac12}^{(1)}(-kx_1)H_{\alpha+\frac12}^{(2)}(k\tau_1)\nonumber,
\end{eqnarray}
where $x_1=\tau_1-\tau_{\text{inf}}$. In the derivation of
these coefficients, we have used an identity of Hankel's function
\cite{handbook}
\begin{equation}
H_{\nu+1}^{(1)}(z)H_{\nu}^{(2)}(z)-H_{\nu+1}^{(2)}(z)H_{\nu}^{(1)}(z)=-\frac{4i}{\pi z}.
\end{equation}

\subsection{Thermal initial condition and the primordial power spectrum of gravitational waves}

{\color{black}Following the discussion in \cite{Gasperini:1993yf},
particle production process in the expanding universe can be
described in terms of Bogoliubov transformations
\cite{2007iqeg.book.....M}. For each mode $k$, the annihilation
and creation operators satisfy the following relations,
\begin{equation}
\hat b_{\bf k}=c_+(k)\hat a_{\bf k}+c_-^*(k)\hat a^{\dagger}_{\bf -k},~~~\hat b_{\bf k}^{\dagger}=c_-(k)\hat a_{\bf -k}+c_+^*(k)\hat a_{\bf k}^{\dagger},
\end{equation}
where $(\hat a_{\bf k},\hat a_{\bf k}^{\dagger})$ are the creation
and annihilation operators for the $|\text{in}\rangle$ state,
while $(\hat b_{\bf k},\hat b_{\bf k}^{\dagger})$ for the
$|\text{out}\rangle$ state. The Bogoliubov coefficients
$c_{\pm}(k)$ depend on the dynamics of the background geometry of
the universe, and satisfy $|c_+|^2-|c_-|^2=1$. In this paper, we
consider the state at the beginning of pre-inflationary stage as
the $|\text{in}\rangle$ state, and the state at the end of
inflationary stage as the $|\text{out}\rangle$ state. If the
$|\text{in}\rangle$ is a vacuum state, i.e. $\hat a_{\bf
k}|\text{in}\rangle=\hat a_{\bf k}|0\rangle=0$, the particle
number in the $|\text{out}\rangle$ is
\begin{equation}
\bar N_k=\langle 0|\hat b_{\bf k}^{\dagger}\hat b_{\bf
k}|0\rangle=|c_-(k)|^2,
\end{equation}
and the corresponding power spectrum can be calculated by
\cite{Baumann:2009ds}
\begin{equation}\label{eq17}
\langle 0| \hat h_{\bf k}\hat h_{\bf
k'}|0\rangle=\frac{4}{a^2(\tau)M_{\text{Pl}}^2}|v_k|^2\delta^3({\bf
k}+{\bf k'}) \equiv \frac{2\pi^2}{k^3}P_h(k)\delta^3({\bf k}+{\bf
k'}),
\end{equation}
where $v_k$ is directly determined by the Bogoliubov coefficient
$c_{-}(k)$, which is given by Eq. (\ref{eq9}).

However, if the $|\text{in}\rangle$ state is a thermal state for
gravitons, i.e. $\hat a_{\bf k}^{\dagger}\hat a_{\bf
k}|\text{in}\rangle=\hat a_{\bf k}^{\dagger}\hat a_{\bf
k}|T\rangle=n_k|T\rangle$ with $n_k=\frac{1}{e^{k/T}-1}$, the
number of gravitons for the $|\text{out}\rangle$ becomes
\cite{Gasperini:1993yf}
\begin{equation}
\bar N_k=\langle T|\hat b_{\bf k}^{\dagger}\hat b_{\bf
k}|T\rangle=|c_-(k)|^2(1+n_k)+n_k(1+|c_-(k)|^2)=|c_-(k)|^2\coth
\left(\frac{k}{2T}\right).
\end{equation}
Here, $T$ denotes the conformal temperature of the gravitons. The
physical temperature $\mathcal{T}$ is given by $\mathcal{T}=T/a$.
In this paper, we always set the present scale factor $a_0=1$, so
the conformal temperature is also the present physical temperature
of the thermal state. Similar to \cite{Gasperini:1993yf}, we have
neglected the numerical factors of order unity in the last step.
The corresponding power spectrum of gravitational waves becomes
\cite{Gasperini:1993yf,bha,zhao2009,das}
\begin{equation}\label{eq17}
\langle T| \hat h_{\bf k}\hat h_{\bf
k'}|T\rangle=\frac{4}{a^2(\tau)M_{\text{Pl}}^2}|v_k|^2 \coth
\left(\frac{k}{2T}\right)\delta^3({\bf k}+{\bf k'}) \equiv
\frac{2\pi^2}{k^3}P_h(k)\delta^3({\bf k}+{\bf k'}),
\end{equation}
which follows,
\begin{equation}
P_h(k)=\frac{2k^3}{\pi^2M^2_{\text{Pl}}a^2(\tau)}|v_k|^2\coth{\left(\frac{k}{2T}\right)}.
\end{equation}
Comparing with the results in Eq. (\ref{eq17}), we find that the
thermal initial state only contributes the extra factor
$\coth({k}/{2T})$ in the power spectrum of PGWs. Using the results
in Eq. (\ref{eq9}), we derive that \cite{das}}
\begin{equation}
P_h(k)=P^{\rm BD}_h(k)|B_1(k)-B_2(k)|^2\coth{\left(\frac{k}{2T}\right)},
\end{equation}
where $P^{\rm BD}_h(k)$ corresponds to the standard power spectrum of PGWs in vacuum presuming Bunch-Davis initial conditions, which are always parameterized as a power-law form. We should notice that the power spectrum in this alternative scenario is just a combination of the standard power spectrum with two additional factors, which are determined by the cosmic evolution of the pre-inflationary stage and the conformal temperature of the thermal state. The modification of the mode functions, due to the presence of the pre-inflationary stage (which has been ignored in the previous works \cite{bha,zhao2009}) leads to the appearance of the first factor in the equation, i.e. $|B_1-B_2|^2$. The second factor, $\coth{\left({k}/{2T}\right)}$, is due to the thermal initial state of PGWs.
Then, the parameterized expression of $P_h(k)$ is given by
\begin{equation}
\label{result}
P_h(k)=A_t\cdot \left(\frac{k}{k_p}\right)^{n_t}\cdot\frac{|B_1(k)-B_2(k)|^2}{|B_1(k_p)-B_2(k_p)|^2}\cdot\frac{\coth{(\frac{k}{2T})}}{\coth{(\frac{k_p}{2T})}},
\end{equation}
where $k_p$ is the pivot scale, $A_t$ is the amplitude of tensor perturbations at $k=k_p$, and $n_t$ is the tensor spectral index, which is zero in the de Sitter inflationary models.

We should mention that, in addition to the PGWs, a thermal spectrum of primordial density perturbations might also exist in the pre-inflationary scenario \cite{bha,das}.  When we use the same parameters used in \cite{das}, we can get the same modification for the primordial power spectrum, since the evolution equations for scalar and tensor perturbations are the same. However, different from PGWs, the nature of the density perturbations depends crucially on the content and the state of matter in the early Universe, in particular, on those of the pre-inflationary period, which are yet to be fully understood. So, similar to our previous work \cite{zhao2009}, we will not consider the density perturbations in this paper, but we generalize our result to a general pre-inflationary stage since we have no idea about Universe before inflation and we should consider the general condition rather that a specific one.

\subsection{Relations between the model parameters}

In order to figure out the complete set of independent parameters in our model, let us investigate the relations between different model parameters. Since the gravitons are massless, the conformal temperature satisfies that $T=a(\tau)\mathcal{T}=\text{constant}$, which is derived from the conservation of entropy in the Universe. As a result, we can infer the relation between conformal temperature in different stages with the scale factor, which can be written as \cite{zhao2009}
\begin{equation}\label{relation_tem}
\frac{\mathcal{T}_0}{\mathcal{T}_i}=\frac{a_i}{a_0}=\frac{a_i}{a_{\text{inf}}}\times \frac{a_{\text{inf}}}{a_p}\times \frac{a_p}{a_0},
\end{equation}
where $\mathcal{T}_i$ is the planck temperature and different subscripts stand for different stages of Universe: ${\rm inf}$ is for the beginning of the inflationary stage, $i$ is
for   beginning of the pre-inflationary stage, $p$ is for the
beginning of the post-inflationary stage and $0$ is for today.
Note that similar to \cite{zhao2009}, we have ignored the
reheating period in this paper \cite{reheat}.

Now, let us evaluate the three factors on the right-hand side of Eq. (\ref{relation_tem}) as a function of cosmological parameters, respectively. Using the Friedmann equation, the first term is given by
\begin{equation}
\frac{a_i}{a_{\text{inf}}}=\left(\frac{\rho_{\text{inf}}}{\rho_i}\right)^{\frac{1}{3(1+w)}}=\left(\frac{M_{\text{inf}}}{M_{\text{Pl}}}\right)^{\frac{4}{3(1+w)}},
\end{equation}
where $M_{\text{inf}}\simeq (\frac{r}{0.01})^{1/4}\times 10^{16}\text{GeV}$ \cite{Baumann:2009ds} is the energy scale of the inflationary stage and $r$ is the tensor-to-scalar ratio. The second term can be converted into the total e-folding number $N$ in the inflationary stage through ${a_{\text{inf}}}/{a_p}=e^{-N}$. The last term can be evaluated by the relation \cite{earlyUniverse}
\begin{equation}
\frac{a_p}{a_0}\simeq \frac{2.73{\rm K}}{M_{\text{inf}}}\left( \frac{3.91}{106.75}\right)^{1/3}.
\end{equation}
Thus, we can get the physical temperature of gravitons today,
\begin{equation}\label{T-N}
\mathcal{T}_0\simeq 6.57\times\left( \frac{r}{2.215\times 10^{10}}\right)^{\frac{1}{3(1+w)}}\times \frac{e^{60-N}}{r^{1/4}}~~~\text{Mpc}^{-1},
\end{equation}
which indicates that the present temperature of thermal gravitational-wave background depends on the model parameters $r$, $w$ and $N$. In the specific case with $w=1/3$, i.e. the pre-inflationary stage is radiation-dominant, the value of $\mathcal{T}_0$ is independent of $r$, which is consistent with the results derived in \cite{zhao2009}. However, in the general pre-inflationary scenario, the value of $\mathcal{T}_0$ significantly depends on the effective EoS $w$. For the reasonable choice $r\sim 0.01$ and $N>60$, if $w=1/3$, we have $\mathcal{T}_0 \lesssim 0.017 \text{Mpc}^{-1}$. However, if $w=0$, it becomes $\mathcal{T}_0 \lesssim 0.0016 \text{Mpc}^{-1}$, and if $w=1$, it is $\mathcal{T}_0 \lesssim 0.18 \text{Mpc}^{-1}$.


Now, let us fix the last parameter $\tau_1$ in Eq. (\ref{a-evolution}).
For the gravitational-wave mode, which corresponds to the size of the horizon when the inflation began, we have $k_{\text{inf}}=a_{\text{inf}}H_{\text{inf}}$, where $H_{\rm inf}$ is the Hubble parameter during inflationary stage. While for the maximum mode we can observe today, we have $k_0=a_0H_0$, where $H_0$ is the Hubble constant today. In order to solve
the horizon problem, we should have $k_{\text{inf}}<k_0$. We
can quantify this relation with e-folding number
\begin{equation}
N(k_0)-N(k_{\text{inf}})=\ln (k_{\text{inf}}/k_0).
\end{equation}
Considering the relation
$\tau_1=\frac{1}{a_{\text{inf}}H_{\text{inf}}}$ (we should note
that there is no minus here due to the definition of $\tau$), we
can get the time when the inflation began

\begin{equation}
\tau_1=\frac{1}{a_0H_0}e^{N-N(k_0)}.
\end{equation}
Being $N(k_0)$ the minimum e-folding number needed to solve the
horizon problem, which can be approximately expressed as
\cite{das}
\begin{equation}
N(k_0)=\ln \left(\frac{a_pH_p}{a_0H_0}\right)=63.59+0.25\times \ln r,
\end{equation}
where $H_p$ is the Hubble parameter just after the inflationary stage. And we can get the e-folding number $N$ from Eq.(\ref{T-N}), which is
\begin{equation}
N=60+\ln{\left[\frac{6.57}{\mathcal{T}_0\times r^{1/4}}\times \left(\frac{r}{2.215\times 10^{10}}\right)^{\frac{1}{3(1+w)}}\right]}
\end{equation}

Using these relations, we can figure out three independent parameters in our model, which are the effective EoS of the pre-inflationary stage $w$, the tensor-to-scale ratio $r$ and the present physical temperature of the thermal gravitons $\mathcal{T}_0$, i.e. the conformal temperature $T$.

\begin{figure}
   \centering
   \includegraphics[width=10cm]{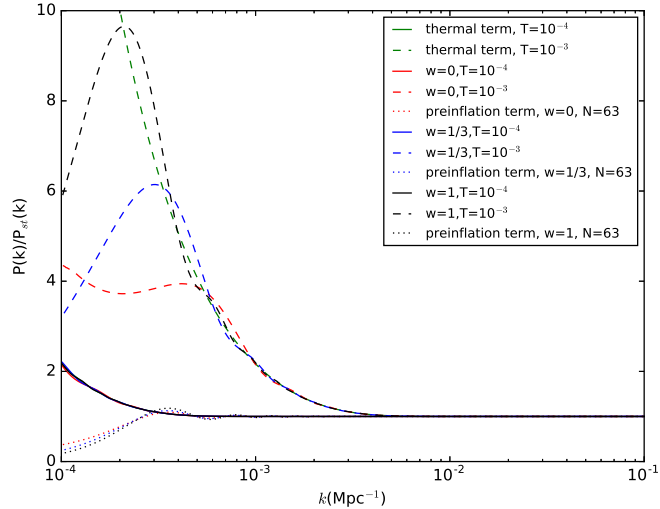}
   \caption{The ratio between power spectrum in the pre-inflationary model and that in standard inflationary scenario with respect to different parameters. {\color{black}The `thermal term' stands for the case where we ignore the third term and only consider the last term in Eq. (\ref{result}), which is contributed by the thermal initial condition. The `preinflation term' stands for the case where we ignore the last term and only consider the third term in Eq. (\ref{result}), which is from the mode mixing in pre-inflationary stage.} Note that, three dashed lines overlap with each other in this figure.}
   \label{f1}
   \end{figure}

We plot the power spectrum ratio, which is the ratio between the power spectra of PGWs in
pre-inflationary model and that in the standard inflationary
scenario, $P_k(k)/P^{\rm BD}_h(k)$ , with respect to different
parameters in Fig. \ref{f1}. To show the different effects brought by the pre-inflationary stage and thermal initial condition respectively, we also plot the power spectrum ratio for the conditions where only one of them is working. We could see that the thermal initial condition will enhance the power spectrum greatly on the large scale and if we have larger $T$, this effect will work on smaller scale, which will be easier to observe it. The modification on the power spectrum caused by pre-inflationary stage is also occurs on large scale, but more complex. It will depress the power spectrum on large scale and bring some wiggles on the scale which is the length of the horizon when the inflation begins. In addition, when $T$ is large enough i.e. $T\geqslant 10^{-3}\text{Mpc}^{-1}$, the difference between different EOS is observable in Fig. \ref{f1}. If $T$ is small, like $T\sim 10^{-4}\text{Mpc}^{-1}$, we could see the effect brought by the thermal initial condition, but it will be difficult the distinguish the different EOS.
At last, we should mention that the necessary to observe these effects is that the inflation did not last too long, which means we have large $T$ or small $N$, otherwise, all these modifications will be washed by the inflationary phase.
It is obvious that the difference
mainly occurs in an extreme large scales, where we should notice
that $k\sim 10^{-4}\text{Mpc}^{-1}$ corresponds to the scale of
our observable Universe. If the inflationary stage lasts long
enough, i.e. the smaller $T$, all the information left by the
pre-inflationary stage will be washed out. However, we could probe
the pre-inflationary stage if the exponential expansion does not
last so long, and the value of $T$ is larger than
$10^{-4}\text{Mpc}^{-1}$.  As we can see from Fig. \ref{f1}, there
are some small wiggles on large but observable scale ($k\sim
10^{-4}\text{Mpc}^{-1}$). These differences will leave signatures
on the B-mode polarization, which we will see in the next section.

\section{The imprint on the CMB B-mode power spectrum and their detection}
The detection the PGWs has attracted
great attention recently. In the low frequency range, the
detection is mainly by observations of the CMB temperature and
polarization fluctuations. In particular, the B-mode polarization
provides a clean information channel for the detection,
which is not contaminated by the density perturbations \cite{B-mode}. In this
paper, we shall only focus on the B-mode polarization, and ignore
the contribution of PGWs in the CMB temperature and E-mode
polarization.

\begin{figure}
   \centering
   \includegraphics[width=10cm]{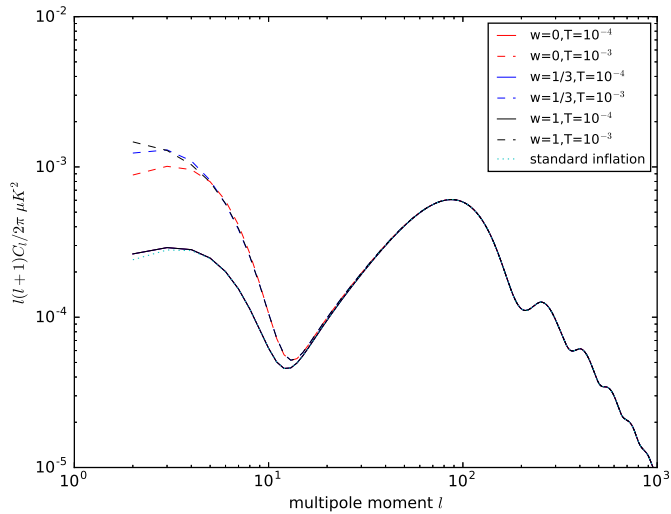}
   \caption{The B-mode power spectra caused by the primordial gravitational waves in the different pre-inflationary models.
   Note that, three dashed lines overlap with each other. In this figure, we have fixed the parameters $r=0.01$ and $n_t=0$.}
   \label{f2}
   \end{figure}

In the linear approximations, the B-mode power spectrum of the CMB
depends linearly on the power spectrum of PGWs. Thus,the
modification of PGWs caused by both the thermal initial state and
pre-inflationary stage can be directly reflected in the B-mode
power spectrum. By employing the power spectrum of PGWs in Fig.
\ref{f1}, in Fig. \ref{f2} we plot the corresponding B-mode power
spectra, which clearly shows its dependence on the
pre-inflationary parameters $w$ and $T$. The main feature of the
figure is the quite large distinction caused by different $T$.
However, for fixed $T$ values, the difference caused by different
$w$ is fairly small. Consistent with Fig. \ref{f1}, we find that,
when $T=10^{-4}$Mpc$^{-1}$, the effect of pre-inflationary stage
is very small, and it is difficult to see the difference between different EOS, which is independent of the effective EoS $w$.
However, if $T=10^{-3}$Mpc$^{-1}$, the effect of pre-inflation is
quite significant in the low multipole range $\ell\lesssim 10$.
For the model with different $w$, the difference is only at the
lowest multipole range $\ell \lesssim 5$, and a larger $w$ always
follows a larger $C_{\ell}^{BB}$.


\subsection{Current constraints}

\begin{figure}
\centering
\subfigure{\includegraphics[width=0.33\textwidth]{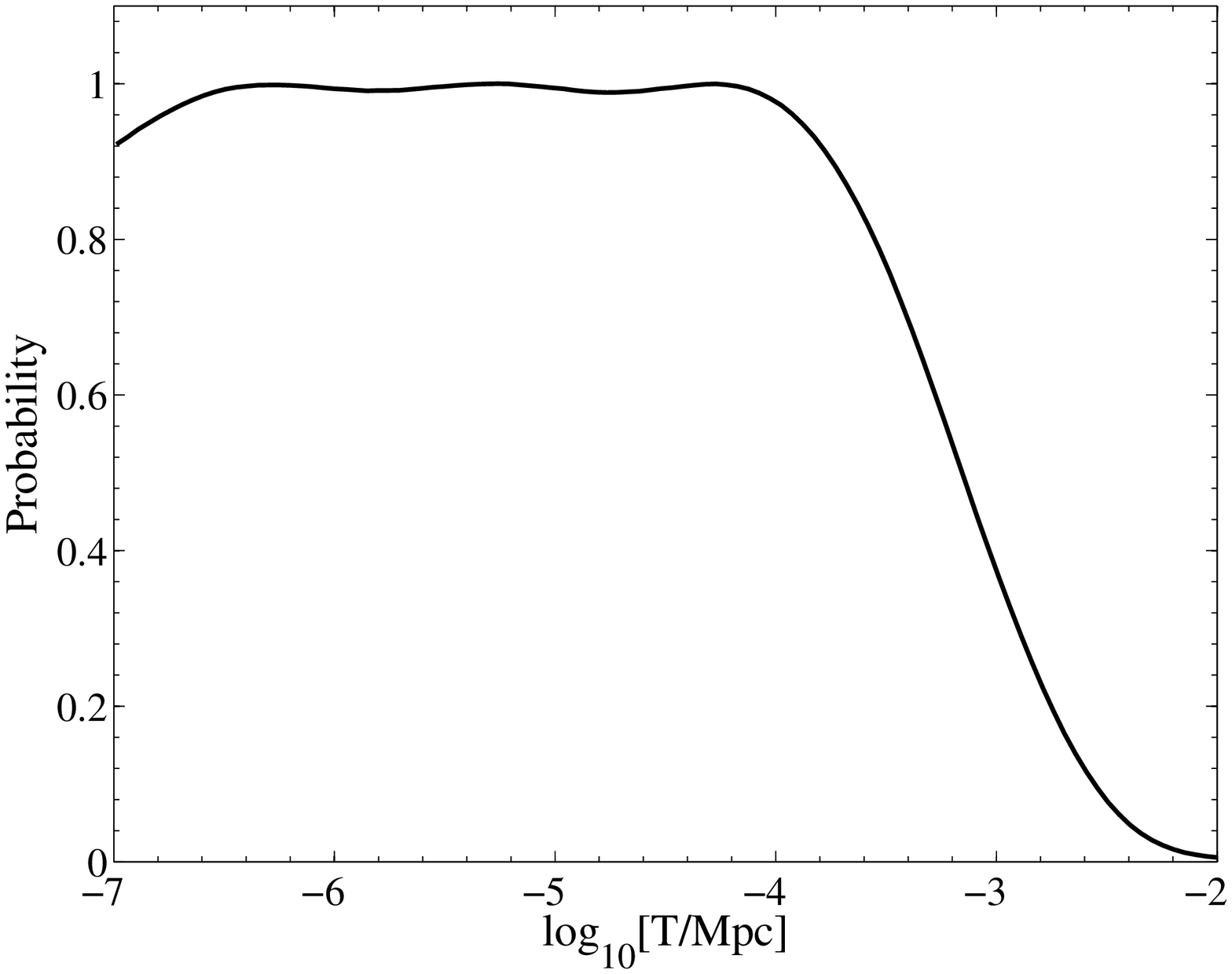}}
\subfigure{\includegraphics
[width=0.33\textwidth]{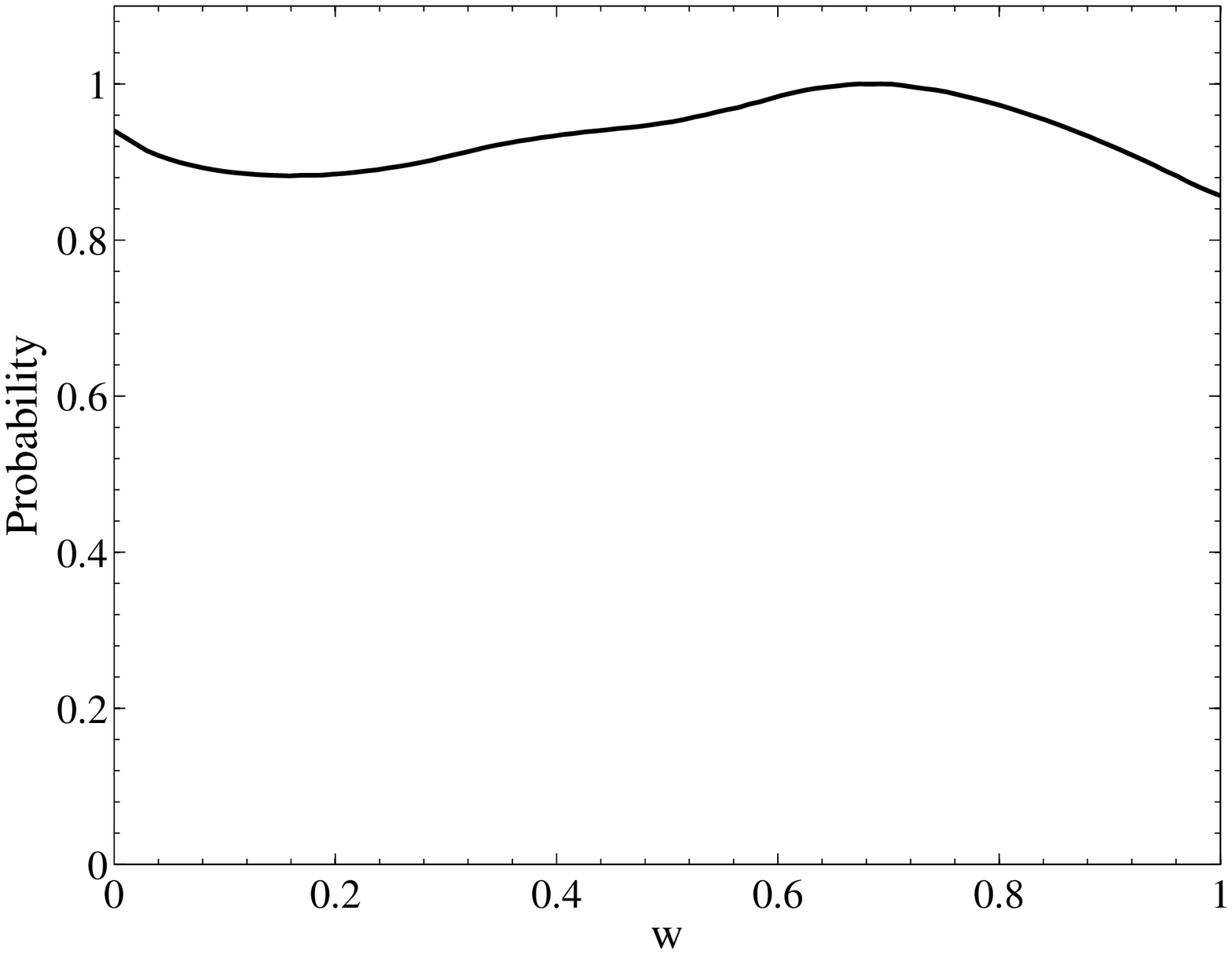}}
\subfigure{\includegraphics
[width=0.33\textwidth]{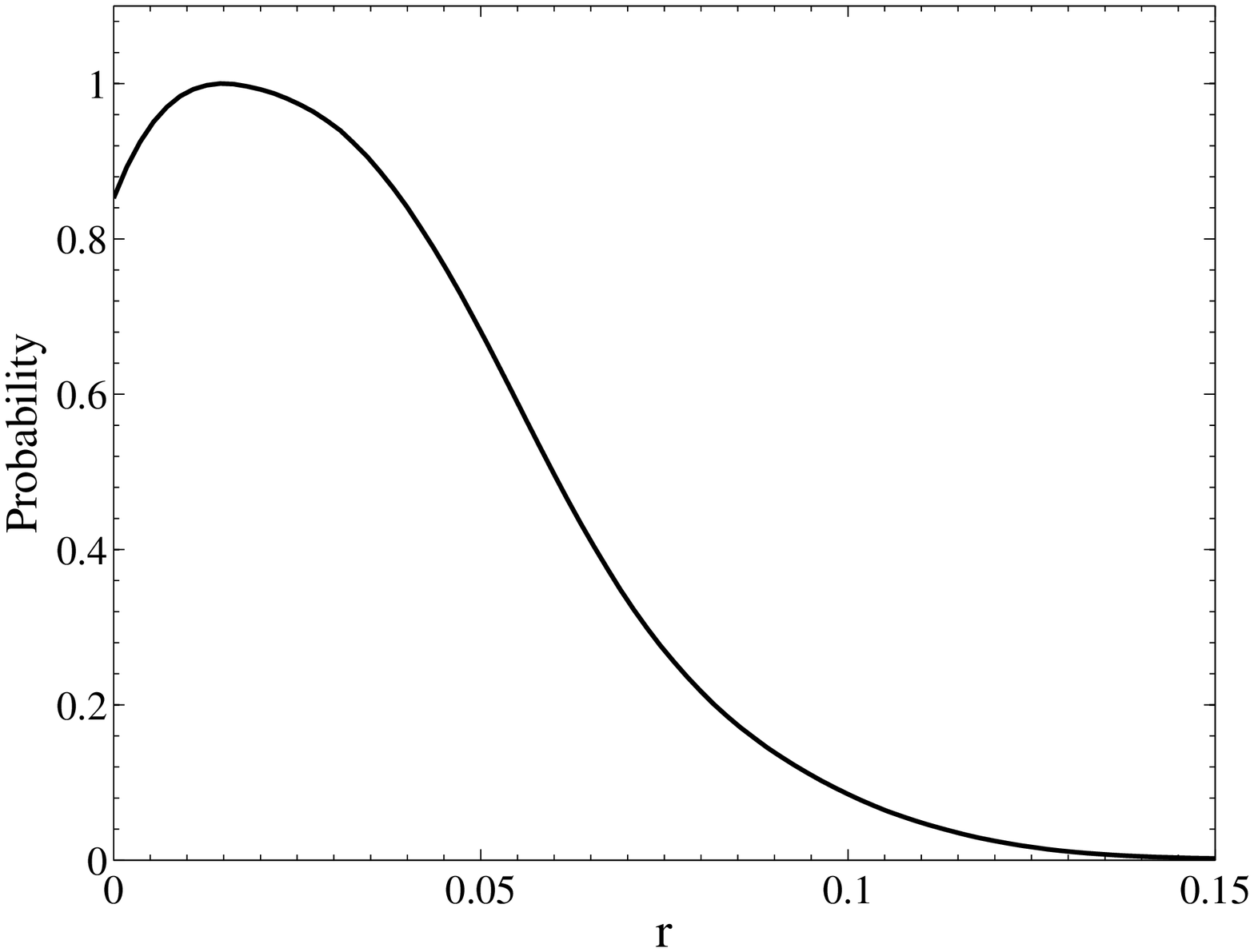}}
\caption{1-dimensional likelihood function for the model
parameters $T$ (upper), $w$ (middle) and $r$ (lower).} \label{f3}
\end{figure}

The modifications of the PGWs and the CMB fluctuations are
expected to be constrained by observations. In this subsection, we
shall perform the CMB likelihood analysis by using the recent CMB
B-mode data, which are derived from the joint analysis of
BICEP2/Keck Array and Planck data \cite{bicep2}. In order to
exclude the influence of density perturbation, we do not consider
the other CMB data, including $TT$, $TE$ and $EE$ power spectra.
For the background cosmological model, we adopted the base
$\Lambda$CDM model with the best-fit parameters derived from the
Planck data $\Omega_{\rm b}h^2=0.02225$, $\Omega_{\rm
c}h^2=0.1198$, $100\theta_{\rm MC}=1.04077$, $\tau_{\rm
reio}=0.079$, $\ln(10^{10}A_s)=3.094$ at $k_0=0.05$Mpc$^{-1}$
\cite{planck-para}. For the PGWs, we fix the spectral index
$n_t=0$, and set free the other parameters $r$, $T$ and $w$. By
applying the modified COSMOMC package, we derive the constraints
on these free parameters. At $95\%$ confident level, they are
given by
\begin{equation} \label{constraint}
r<0.084,~~T<5.01\times10^{-4}{\rm Mpc}^{-1},
\end{equation}
being the constraint on $w$ too loose. The 1-dimensional
likelihood for them are given in Fig. \ref{f3}.

The upper limits on the conformal temperature $T$ of the thermal gravitational-wave background, and the tensor-to-scalar ratio allow us to place interesting constraints on the physics of inflationary era. Employing the relation in Eq. (\ref{T-N}) and the results in Eq. (\ref{constraint}), we get the constraint on the total e-folding number $N>63.5$ for the model with effective EoS $w=1/3$. While for the model with effective EoS $w=1$, this bound becomes $N>65.7$. Both bounds are consistent with the e-folds parameter required to solve the flatness, horizon and monopole problems in the standard hot big-bang cosmological model \cite{earlyUniverse,Baumann:2009ds}.

\subsection{Forecast for the potential CMBPol detections}
Although the B-mode polarization in CMB induced by PGWs has not
yet been detected due to present instrument ability and the
various contaminations \cite{no-detection}, it is expected that in
the near future, the detection abilities of various ground-based,
balloon-borne and space-based experiments will be greatly
improved. As we noted above, the features of the pre-inflationary
period are mainly on the large scales, which is more probable to
be observed by future CMB satellites. In this section, we will use
Fisher information matrix to forecast for CMBPol mission
\cite{cmbpol,zhao-cmbpol}, a typical CMB satellite of next
generation.

Similar to the discussion above, here we only consider the B-mode information channel, so the
Fisher information matrix can be written as \cite{Tegmark:1996bz}
\begin{equation}
{\bf F}_{ij}=\sum\limits_{\ell} \frac{\partial
C_{\ell}^{BB}}{\partial p_i}\frac{\partial C_{\ell}^{BB}}{\partial
p_j}\frac{1}{(\Delta D_{\ell}^{BB})^2},
\end{equation}
where $p_i~(i=1,2,3,...)$ are the model parameters. In our investigation, they are conformal temperature for gravitons $T$, effective EoS of pre-inflationary stage $w$ and the
tensor-to-scalar ratio $r$. The quantity $\Delta D_{\ell}^{BB}$ is
the standard deviation of the estimator $D_l^{BB}$, which is given by
\begin{equation}
\Delta
D_{\ell}^{BB}=\sqrt{\frac{2}{(2\ell+1)f_{\text{sky}}}}(C_{\ell}^{BB}+N_{\ell}^{BB})
\end{equation}
where $f_{\text{sky}}=0.8$ is the sky-cut factor for the CMBPol mission. $C_{\ell}^{BB}$ and
$N_{\ell}^{BB}$ correspond to theoretical B-mode and noise power
spectra, respectively.

\begin{table}
\centering
\begin{tabular}{|c|c|c|c|c|c|}
\hline Frequency [GHz] & 45 & 70 & 100 & 150 & 220\\
\hline $\theta_F$ [arcmin] & 17 & 11 & 8 & 5 & 3.5\\
\hline $\Delta_T$ [$\mu$K-arcmin] & 5.85 & 2.96 & 2.29 & 2.21 & 3.39\\
\hline
\end{tabular}
\caption{Experimental specifications for the CMBPol (EPIC-2m) mission. \cite{cmbpol,zhao-cmbpol}}
\label{T_experiment}
\end{table}

The noise arises from three types of sources. The first one corresponds to the noise on the instrument. For a single frequency channel, the power
spectrum (after deconvolution of the beam window function) is
given by \cite{zhao-cmbpol}
\begin{equation}
N^{\text{ins}}_{\ell}(\nu)=2\cdot (\Delta_T)^2
\exp\left[\frac{\ell(\ell+1)\theta_F^2}{8\ln 2}\right],
\end{equation}
where $\Delta_T$ is the noise for temperature and $\theta_F$ is
the full width at half maximum (FWHM) (their values for different frequency
bands can be found in Table \ref{T_experiment}).

\begin{table}
\centering
\begin{tabular}{|c|c|c|c|c|c|c|}
\hline Parameters & $A_{X}~[\mu {\rm K}^2]$ & $\nu_X$ [GHz] & $\ell_{X}$ & $\alpha_X$ & $\beta_X$ & T [K]\\
\hline Synchrotron & $2.1\times 10^{-5}$ & 65 & 80 & -2.6 & -2.9 & ---\\
\hline Dust emission & 0.169 & 353 & 80 & -2.42 & 1.59 & 19.6\\
\hline
\end{tabular}
\caption{A list of foreground parameters for $f_{\text{sky}}=0.8$,
where $X$ denotes $S$ (synchrotron) or $D$ (dust emission).
\cite{Huang:2015gca,Creminelli:2015oda}}
\label{T_foreground}
\end{table}

The second contamination is the polarized foreground, which comes from free-free, synchrotron, dust emission and the extra-galactic sources such as radio sources and dusty galaxies. However, only the synchrotron and the thermal dust emission are important in the frequency range of CMBPol. Their power spectra in thermodynamical temperature are \cite{Huang:2015gca,Creminelli:2015oda,Escudero:2015wba}
\begin{align}
C_{\ell}^{S}(\nu)&=\frac{2\pi}{\ell(\ell+1)}~ A_S~(W_{\nu}^{S})^2~\left(\frac{\ell}{\ell_S}\right)^{\alpha_S}, \\
C_{\ell}^{D}(\nu)&=\frac{2\pi}{\ell(\ell+1)}~A_D(W_{\nu}^{D})^2~\left(\frac{\ell}{\ell_D}\right)^{\alpha_D},
\end{align}
where the parameters can be found in Table \ref{T_foreground}. $W_{\nu}^S$ and $W_{\nu}^D$ are defined by
\begin{eqnarray}
W_{\nu}^S&=&\frac{W_{\nu_S}^{\text{CMB}}}{W_{\nu}^{\text{CMB}}}\left(\frac{\nu}{\nu_S}\right)^{\beta_S}, \nonumber \\
W_{\nu}^S&=&\frac{W_{\nu_D}^{\text{CMB}}}{W_{\nu}^{\text{CMB}}}\left(\frac{\nu}{\nu_D}\right)^{1+\beta_D}\frac{\exp(h\nu_D/k_BT)-1}{\exp(h\nu/k_BT)-1},
\nonumber
\end{eqnarray}
where
\begin{equation}
W_{\nu}^{\text{CMB}}=\frac{x^2e^x}{(e^x-1)^2},~~~~x=\frac{h
\nu}{k_BT_{\text{CMB}}}.
\end{equation}
In this paper, we will assume a residual factor $\sigma^{X}=0.01~(X=S,D)$ to be responsible for the subtraction level rather than discuss the details of the subtraction process, where $X=S,D$ indicates synchrotron and dust emission, respectively \cite{cmbpol,zhao-cmbpol,Huang:2015gca,Creminelli:2015oda,Escudero:2015wba}.

The third contamination for the primordial B-mode polarization is caused by the cosmic weak lensing effect. If considering the secondary effect, the E-mode polarization can induce the B-mode polarization along the line-of-sight between the observer and the last-scattering surface through cosmic weak lensing \cite{weak-lensing}. We also use a residual factor $\sigma^{\text{lens}}$ to describe the subtraction level. In this paper, we consider the case with $\sigma^{\text{lens}}=1$, which means that we do not consider the subtraction of the weak lensing effect by any de-lensing techniques.

Since the CMB experiments have several frequency channels, the optimal channel combination will give the total noise power spectrum, which can be written as  \cite{zhao-cmbpol,Escudero:2015wba}
\begin{equation}
N_{\ell}^{BB}=N^{\text{eff}}_{\ell}+C^{\text{lens}}_{\ell}\cdot
\sigma^{\text{lens}},
\end{equation}
where
\begin{align}
(N_{\ell}^{\text{eff}})^{-2} & =\sum\limits_{i,j\ge
i}^{N_{\text{chan}}}[(R_{\ell}^F(\nu_i)+N^{\text{ins}}_{\ell}(\nu_i))
\nonumber
\\
& \times(R_{\ell}^F(\nu_j)+N^{\text{ins}}_{\ell}(\nu_j))\times \frac12 (1+\delta_{ij})]^{-1}, \\
R_{\ell}^F(\nu) & =\sum\limits_{X=S,D}\left [
\sigma^X(\nu)C_{\ell}^X(\nu) \right. \nonumber
\\
& \left.
+N^{\text{ins}}_{\ell}(\nu)\frac{4}{N_{\text{chan}}(N_{\text{chan}}-1)}\frac{C_{\ell}^X(\nu)}{C_{\ell}^X(\nu_F)}
\right],
\end{align}
where $i,j$ correspond to different frequency channels,
$N_{\text{chan}}$ is the number of detection channels and $\nu_F$
is the highest and lowest frequency channel included in the
cosmological analysis for dust and synchrotron respectively, i.e.
that listed in Table \ref{T_foreground}.

By calculating the Fisher information matrix, we can get the uncertainties of the
model parameters with Cramer-Rao bound, which is
\cite{Tegmark:1996bz}
\begin{equation}\label{delta-T}
(\Delta {\hat p_i})^2=\langle (\hat p_i-\langle \hat p_i\rangle)^2 \rangle={\bf F}^{-1}_{ii}.
\end{equation}

\begin{figure}
   \centering
   \includegraphics[width=10cm]{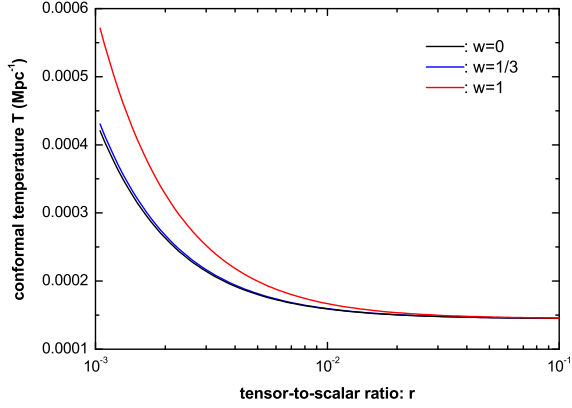}
   \caption{Smallest value of conformal temperature $T$ which can be detected by CMBPol.}
   \label{f4}
   \end{figure}

In order to detect the feature of the pre-inflationary stage, we should focus on the determination of the conformal temperature $T$,
which is equivalent to the total e-folding number $N$ through Eq. (\ref{T-N}). A larger e-folding number $N$, i.e. a longer inflationary stage,
always follows a smaller $T$ value. Using expression (\ref{delta-T}), we calculate the value of $\Delta T$ for given values of parameters $T$, $w$ and $r$.
Similar to the previous work \cite{zhao2009}, for a fixed values for $r$ and $w$, we quantify the smallest measurable value for $T$, which is
determined by the condition $T=\Delta T$. In Fig. \ref{f4}, we plot the lowest bounds of $T$ for various $w$ and $r$. Consistent with the previous
work \cite{zhao2009}, for the case with larger $r$, i.e. $r>0.01$, the attainable limit on the conformal temperature is $T\ge 1.5\times 10^{-4}$Mpc$^{-1}$,
which is independent of the effective EoS $w$. Employing the relation in (\ref{T-N}), we find that for the case with $r>0.01$ and $w=1/3$, a potential
detection of thermal gravitational-wave background would indicate that the number of total e-folds is $N<64.7$. However, in the case with smaller $r$,
the limit value of $T$ becomes larger as well. In addition, the effect of $w$ become significant. For the models with fixed $r=0.001$, we have the
detection limit of $T=4.2\times10^{-4}$Mpc$^{-1}$ for $w=0$. While it becomes $T=4.3\times10^{-4}$Mpc$^{-1}$ for the model with $w=1/3$,
and $T=5.7\times10^{-4}$Mpc$^{-1}$ for the model with $w=1$. Thus, for the model with $r=0.001$, a potential detection of thermal background
would show that $N<61.1$ for $w=0$, $N<63.7$ for $w=1/3$ and $N<66.0$ for $w=1$.

In order to investigate whether it is possible to constrain the
effective EoS $w$ of the pre-inflationary period, we calculate the
value $\Delta w$ by using the formula in Eq. (\ref{delta-T}). Even
in the most optimistic case with $r=0.1$ and $T=0.0005$Mpc$^{-1}$,
we get $\Delta w=0.53$ for the model with $w=1/3$ and $\Delta
w=2.06$ for the model with $w=1$. So, it seems impossible to well
determinate the value of effective EoS by the potential
observations of CMBPol mission.


\section{Discussions and conclusions}

 Understanding the expansion history of the Universe is one of the fundamental tasks for the modern cosmology. Recent work \cite{das} shows that a thermal initial condition and a pre-inflationary stage will modify the primordial power spectrum of scalr perturbation and the CMB temperature power spectrum. With respect to the density perturbation, which will bring the complexity for the coupling with the matter, primordial gravitational waves only depends on the expansion history of Universe, which is a unique way to probe Universe from the pre-inflationary era to the present stage. Previous work \cite{zhao2009} just study the impact on the primordial power spectrum of tensor perturbation and CMB B-mode brought by the thermal initial condition, which ignored the evolution of PGWs in the pre-inflationary stage. Primordial gravitational waves,
 generated in the early Universe, provide the unique way to probe it from the pre-inflationary era to the present stage. In this paper, we studied the
 evolution of PGWs in a general scenario, where a general pre-inflationary epoch precedes the usual inflationary stage.
 Comparing the PGWs in the usual inflationary model with Bunch-Davies vacuum as the initial condition, we found that the power
 spectra are modified in two aspects: One is the mixture of the perturbation modes, which is caused by the presence of
 the pre-inflationary stage, and the other is the thermal initial state, which is formed when the gravitons decoupled from the
 thermal equilibrium with other components in the Planck epoch. The current power spectrum of PGWs encodes the information of
 initial state of the Universe, the effective EoS of pre-inflationary period, as well as the physics of inflationary stage.
 So, its detection may shed light onto quantum gravity effects, which become important at Planck energy scale. With reasonable assumption,
 we found that this spectrum is quantified by three parameters, the conformal temperature of the thermal gravitational-wave background $T$,
 the effective EoS of pre-inflation $w$, and the tensor-to-scalar ratio $r$. In addition, since the similarity of the evolution of the scalar and tensor perturbation, our result is consisdent with \cite{das}.

 Using the recent CMB B-mode observations by BICEP2/Keck Array and Planck satellite, we derived that,
 at $95\%$ confident level, $r<0.084$ and $T<5.01\times10^{-4}$Mpc$^{-1}$. These upper limits allow us to
 place interesting constraints on the total number of e-folds $N$ of inflationary era, which is $N>63.5$
 for the model with $w=1/3$, and $N>65.7$ for the model with $w=1$. Moreover, considering the noise levels
 and the foreground emission, we also forecast the detection ability of the future CMBPol mission, and found that if $r>0.01$,
 the detection is possible as long as $T>1.5\times10^{-4}$Mpc$^{-1}$, which is independent of the value of $w$.
 However,if $r$ is small, the detection becomes more difficult: For the fiducial mode with $r=0.001$ and $w=1/3$,
 we need $T>4.3\times10^{-4}$Mpc$^{-1}$ for the detection, and for the model with the same $r$ but $w=1$, it becomes $T>5.7\times10^{-4}$Mpc$^{-1}$
 for the possible detection. On the other hand, an absence of observational evidence for a thermal background would indicate one of
 the two possibilities: Either the initial state of gravitational-wave background was not thermal, or alternatively,
 that the number of e-folds is too large so that the present day conformal temperature is redshifted to be too small.

\section*{Acknowledgements}
We appreciate the helpful discussion with Yi-Fu Cai and Dong-Gang Wang. This work is supported by NSFC No. J1310021, 11603020, 11633001, 11653002, 11173021, 11322324, 11421303, project of Knowledge Innovation Program of Chinese Academy of Science and the Fundamental Research Funds for the Center Universities. Our data analysis made the use of CAMB \cite{camb} and COSMOMC \cite{cosmomc}.

\end{document}